\documentclass[a4paper,11pt]{article}
\usepackage{pos}
\usepackage{subcaption}
\title{A novel approach to identify blazar emission states using clustering algorithms}
 \ShortTitle{Clustering algorithms to identify blazar emission states}

\author*[a,b]{L.~Heckmann}
\author[a]{D.~Paneque}
\author[b]{A.~Reimer}

\affiliation[a]{Max-Planck-Institut f\"ur Physik, D-80805 M\"unchen, Germany}
\affiliation[b]{Universität Innsbruck, Institut für Astro- und Teilchenphysik, A-6020 Innsbruck, Austria}

\emailAdd{heckmann@mpp.mpg.de}

\abstract{Even after decades of multi-wavelength (MWL) observations, blazars still remain mysterious objects. Their extreme variability and variety of emission characteristics observed during different time periods make it hard to understand the fundamental processes behind their emission. Thus, a robust identification and characterization of the different emission states among blazars is vital to investigate the underlying processes causing the observed emission.
In this contribution, we present a novel technique to determine emission states across MWL lightcurves (LCs) of blazars using a clustering algorithm. Using the Extreme Deconvolution algorithm, we apply a Gaussian Mixture model to the 12-year long-term LC of one of our archetypal
blazars, Mrk 501. The two main advantages of the method are that, compared to more conventional methods, such as the Bayesian block algorithm, it considers multiple wavebands simultaneously and it is not dependent on the order in time of the data points. This allows to assign data points to the same emission state even though they are separated by other states in time.
The well sampled gamma-ray, X-ray and radio LCs used as input allow to identify six clusters. The clustering is mainly driven by the X-ray flux, showing different levels of quiescent, intermediate and high flux states. However, the radio flux reveals a more complicated pattern,
dividing some of the X-ray flux levels in low and high-radio flux states. This suggests that multiple emission regions maybe responsible for the radio to gamma-ray flux.}

\ConferenceLogo{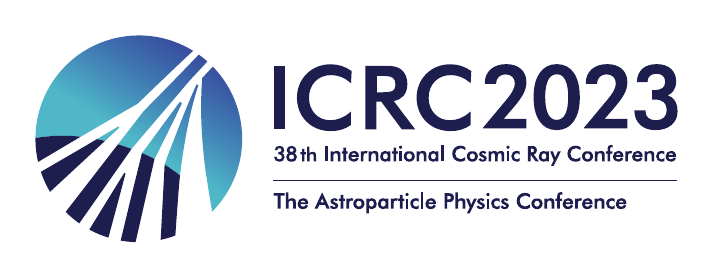}

\FullConference{%
38th International Cosmic Ray Conference (ICRC2023)\\
  26 July - 3 August, 2023\\
  Nagoya, Japan}

\begin{document}
\maketitle

\section{Introduction}

Blazars are among the most energetic sources in our Universe, which accelerate particles to highly relativistic energies. However, even though they have been observed for decades, their underlying emission mechanisms are far from understood. Their extremely variable behaviors, both in flux and spectral characteristics, make it hard to understand the fundamental processes behind the emission. Thus, a robust identification and characterization of their various emission states is vital to enhance our understanding behind blazar emission. 

Most current approach of investigating blazar data concentrates on selecting certain interesting time epochs (often flares) of a single blazar and use it to explore its properties in detail. However, for these limited time intervals and emission states, a variety of theoretical scenarios are usually able to explain the observations, limiting our
capability of inferring how the observed emission is produced. While well sampled and long-term MWL observations are necessary to break this degeneracy, they are currently limited to only a handful of blazars over the whole electromagnetic spectrum up to the highest energies.
But new instruments, such as the Cherenkov Telescope Array (CTA) in the very-high-energy gamma-ray (VHE) regime, and growing numbers of MWL monitoring campaigns will rapidly extend the available data sets. As the available data grows, new statistical methods are necessary to ensure that we can exploit the full potential of both the current and future MWL and multi-messenger data.

In this contribution, we propose a novel technique that could exploit the full potential of  MWL LCs. Cluster algorithms offer the possibility to identify blazar emission states in an unbiased way taking into account all of the available wavebands at the same time while being independent of the order in time of the individual data points. 
We present a first feasibility study using the 12-year MWL LC of one of our archetypal blazars, Mrk 501, and a Gaussian Mixture Model (GMM) as a clustering tool. The following sections present the employed algorithm, its application to the data set, and first interpretations of the results.

\section{The algorithm}
Cluster algorithms are commonly used to identify specific groups, clusters, in a data set. While members of the same cluster should be as similar as possible, members of different clusters should be as different as possible and the definition of similarity and dissimilarity has to be clearly defined \cite{Jain_1988, Cluster_Springer}. Nowadays, a large variety of clustering algorithms exist, such as K-means \cite{kmeans}, GMM \cite{GMM} or DBSCAN \cite{DBScan} with most of them having implementations in the scikit-learn\footnote{\url{https://scikit-learn.org/stable/modules/clustering}} python package.

However, almost all of these wide-spread algorithms do not allow the assignment of inhomogeneous measurement errors on the input data which, for astrophysical data, especially involving different wavebands and instrument, is vital. An exception is the extreme deconvolution method \cite{xdgmm}, which generalizes the GMM approach to include covariances on the input data. It has already been used on astrophysical data to identify pulsar glitch amplitudes by \cite{Reddy_2022}, or to classify gamma-ray bursts based on their duration and hardness \cite{Bhave_2022}. In addition, it is used as a prediction tool, e.g. for anticipating supernova parameters based on the host galaxy parameters \cite{Holoien_2017}, or to identify high-redshift quasars for spectroscopic follow up by \cite{Nanni_2022}.

A GMM with $k$ components, $k$ clusters, is defined by a combination of $k$ multi-variant Gaussians each defined by their specified means $\mu_i$ and covariance matrices $V_j$. The mixing strength of each of the $k$ components is given by the mixing proportions $\alpha_i$. The extreme deconvolution mechanism assumes that the observations are a noisy projection of the true values assuming Gaussian noise with a known covariance matrix. To find the best fit model and therefore clustering for the data, one has to maximize the likelihood using this assumptions, which itself is a Gaussian mixture. An expectation–maximization algorithm
is used as an iterative process to ease this process. The iterative process continues until a certain minimum threshold, tolerance, in the improvement of the likelihood is met, or if a maximum number of iterations is exceeded. For more details see \cite{GMM, xdgmm}.

\subsection{Application to the data}

To apply the extreme deconvolution to blazar LCs, we used the XDGMM\footnote{\url{https://www.astroml.org/modules/generated/astroML.density_estimation.XDGMM.html}} implementation in the astroML \cite{astroML} python package, which takes the data points in combination with their covariance matrices as an input. We chose a maximum number of iteration steps of $10^5$, which is large enough for the algorithm to reach the tolerance level when applied to our data set, and gives stable results. As tolerance, we kept the default value of $10^{-5}$.

To choose a suited data set for this feasibility study, two main factors are considered: maximizing the information, which means maximizing the number of input dimensions and number of data points, and minimizing the introduced bias. For blazars, most data sets are concentrated on flaring episodes and only sparse data are available, which could introduce biases towards different emission states. However, the blazar Mrk\,501 has been monitored continuously since 2008 and owing to its proximity and therefore brightness, it is detected even during its low-activity states up to the VHE. Hence, we chose the long-term MWL LC of Mrk\,501 from \cite{2023ApJS..266...37A} as a data set to investigate the potential of cluster algorithms to identify different emission states. 

To use the MWL LCs as input data for the clustering, we adopt the 14-day binning of the \textit{Fermi}-LAT LC for all other wavebands using the weighted mean of the flux points. Only bins with simultaneous data in all wavebands are selected, which is why we excluded the VHE data for the analysis. The VHE flux measurements are far more sparse and would significantly reduce the number of data points available, and additional systematic errors due to the different instruments and analysis settings in the LC would need to be added. Additionally, only the 2-10\,keV LC of the X-ray data is used for this first study to avoid biasing the results by including both X-ray LCs, which are two very similar data sets. This means, our input data set has three dimensions, the flux values of \textit{Fermi}-LAT, \textit{Swift}-XRT and OVRO.
As the first step of the analysis, all input parameters are scaled to have the same mean value and standard deviation to circumvent biases of different scales in the data set dominating the clustering. We chose a mean value of 10 and standard deviation of 1, and scaled each LC to these values.

To choose the optimal number of clusters for the algorithm, both the Akaike Information Criterion (AIC) \cite{AIC} and Bayesian Information Criterion (BIC) \cite{BIC} are employed. Both of the criteria compare the preference of one models versus another for non-nested models by evaluating the improvement in the likelihood of the model while penalizing for higher degrees of freedom. While the BIC penalizes in a stricter way than the AIC as soon as the number of data points increases above 7, the AIC is a safer choice for cases with a small number of samples, since it tends to under-fit less than the BIC. The opposite is the case for higher number of dimension \cite{Acquah_2010}. Since the number of free parameters will change with future data sets, we employ both criteria in this study. However, the AIC will most probably give results better adjusted to the small data sample for now.

Owing to the random initialization of the algorithm, different optimized results can be reached for different runs of the algorithm, especially when using the less stable AIC. To nonetheless achieve a stable and reproducible result, we ran the algorithm 100 times for each possible number of clusters (1 to 10) and computed the AIC and BIC. We then identified the number of clusters most frequently associated with the dominating minimum criterion value.
To check the significance and that the obtained clustering is not caused by random fluctuations in the LCs, 10,000 uncorrelated LCs are simulated for each waveband using the same method as prescribed in \cite{2023ApJS..266...37A}. We then applied the same clustering procedure to the simulated LCs and computed their optimal number of clusters to compare with the real data.

Finally, the resulting models from the data are used to assign each data point to a certain cluster and the corresponding potential emission state. For each data point, the cluster with the highest probability when evaluating its probability density function at this point is chosen.

\section{The results}

We applied the XDGMM algorithm 100 times for a number of clusters between 1 to 10 to the three-dimensional input data set. Table~\ref{tab:cluster_real} summarizes the occurrence of selected number of clusters among the 100 runs of the algorithm for the two information criteria.

\def\boxit#1{%
\smash{\color{blue}\fboxsep=0pt\llap{\rlap{\fbox{\strut\makebox(#1,28pt){}}}~}}\ignorespaces
}
\setlength{\tabcolsep}{0.5em}
\begin{table}[h]
\centering
\begin{tabular}{  c | c c  c c  c c c  c c c}     % 9 columns 
 Nr of clusters & 1 & 2 & 3 & 4 & 5 & 6 & 7 & 8 & 9 & 10\\
\hline\hline
BIC  & - & - &  \textcolor{blue}{100} & - & - & - & - & - & - & - \\
\hline
AIC  & - & - & -  & 14 & 22 & \textcolor{blue}{47} & 17 & - & - &  - \\
\end{tabular}
\caption{Summary of preferred number of clusters for the three-dimensional data sets for 1 to 10 clusters. Shown are the occurrences of optimal number of clusters obtained for the 100 clustering runs determined by the minimum BIC or AIC applied to the data. The most frequently obtained number of clusters from the data are marked in blue.} 
\label{tab:cluster_real}
\end{table}

Using the BIC, we obtained an optimized number of clusters of 3, which is chosen 100 out of 100 times. The outcome of applying the same procedure to 100 simulations is shown in Table~\ref{tab:cluster_sim}. An optimal number of 3 clusters is reached in 0.06\% of the cases, while most frequently (95.66\%) 1 cluster is chosen for the simulations.

\def\boxit#1{%
\smash{\color{blue}\fboxsep=0pt\llap{\rlap{\fbox{\strut\makebox(#1,28pt){}}}~}}\ignorespaces
}
\setlength{\tabcolsep}{0.5em}
\begin{table}[h]
\centering
\begin{tabular}{  c | c c  c c  c c c  c c c}     % 9 columns 
 Nr of clusters & 1 & 2 & 3 & 4 & 5 & 6 & 7 & 8 & 9 & 10\\
\hline\hline
BIC  & 9,566 & 428 &  \textcolor{blue}{6} & - & - & - & - & - & - & - \\
\hline
AIC  & 1,562 & 2,695 & 2,566  & 1,832 & 895 & \textcolor{blue}{339} & 88 & 23 & - &  - \\
\end{tabular}
\caption{Summary of selected number of clusters for the 10,000 simulations for 1 to 10 clusters. The number corresponding to the most frequently obtained number of clusters from the data is marked in blue.} 
\label{tab:cluster_sim}
\end{table}

For the AIC, 6 clusters were obtained as an optimal number of clusters most frequently (47 times out of 100). Comparing it with the simulations, we reached an optimal number of 6 cluster in 3.4\% of the 10,000 cases, while the most commonly obtained value in the simulation sample is 2 clusters, which is reached in 26.95 of the cases.

\begin{figure}[h] 
     \centering
     \begin{subfigure}{\textwidth}
    \centering
    \includegraphics[width=\textwidth]{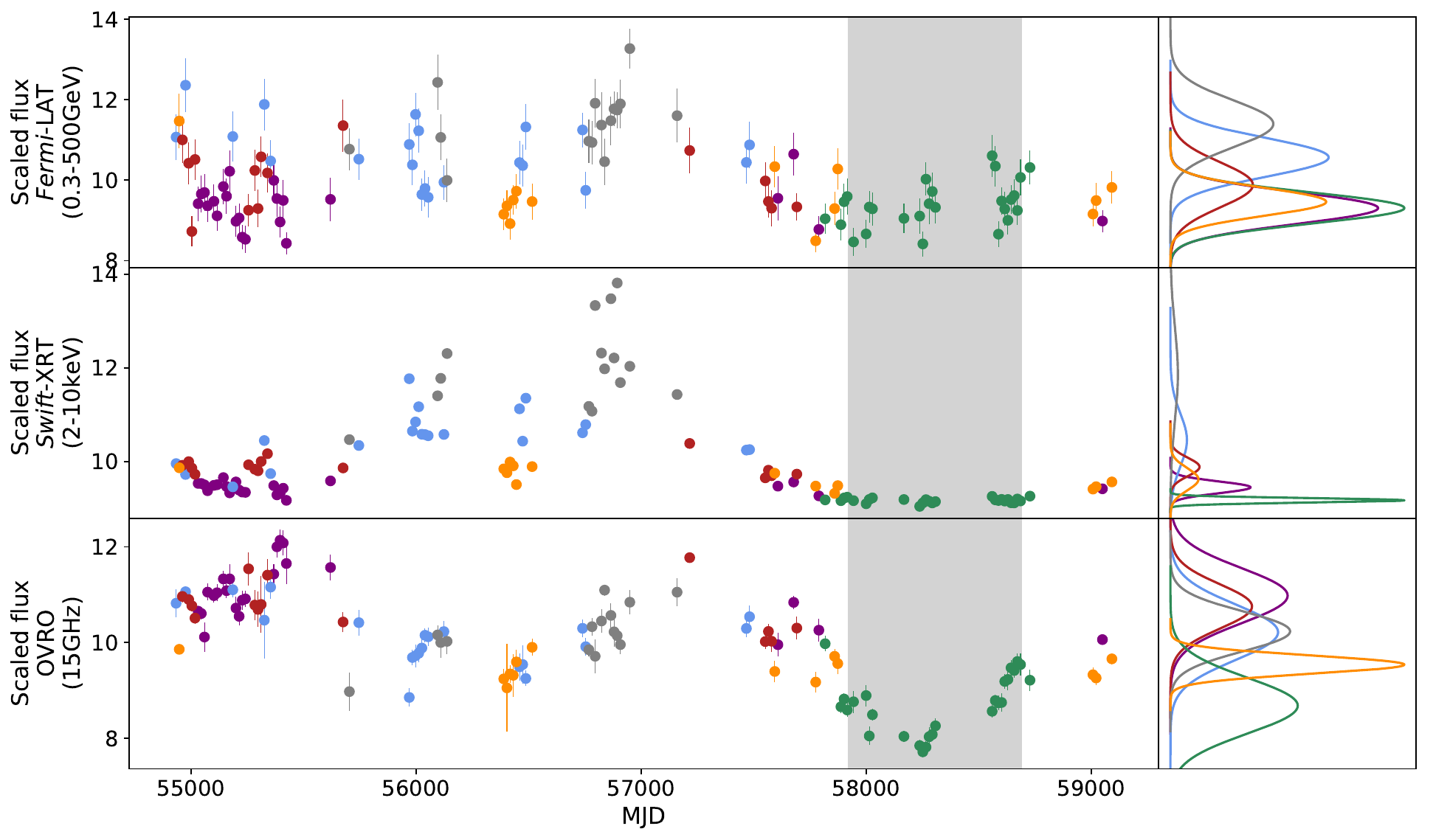}
    \caption{LCs used for the clustering with color schemes assigned to mark which points are assigned to which cluster. The low-state period of Mrk\,501 from \cite{2023ApJS..266...37A} is shown for comparison in grey. The right panels show the final Gaussian Mixture models used for the clustering. }
    \label{fig:lc_cl_of_cl6}
      \vspace*{-50ex}  % Tune this to the image height.
      \hspace{-5cm} \textcolor{lightgray}{\rotatebox{25}{\scalebox{2}{PRELIMINARY}}}
    \vspace*{36ex}
    \end{subfigure}
     \begin{subfigure}{\textwidth}
    \centering
    \includegraphics[width=\textwidth]{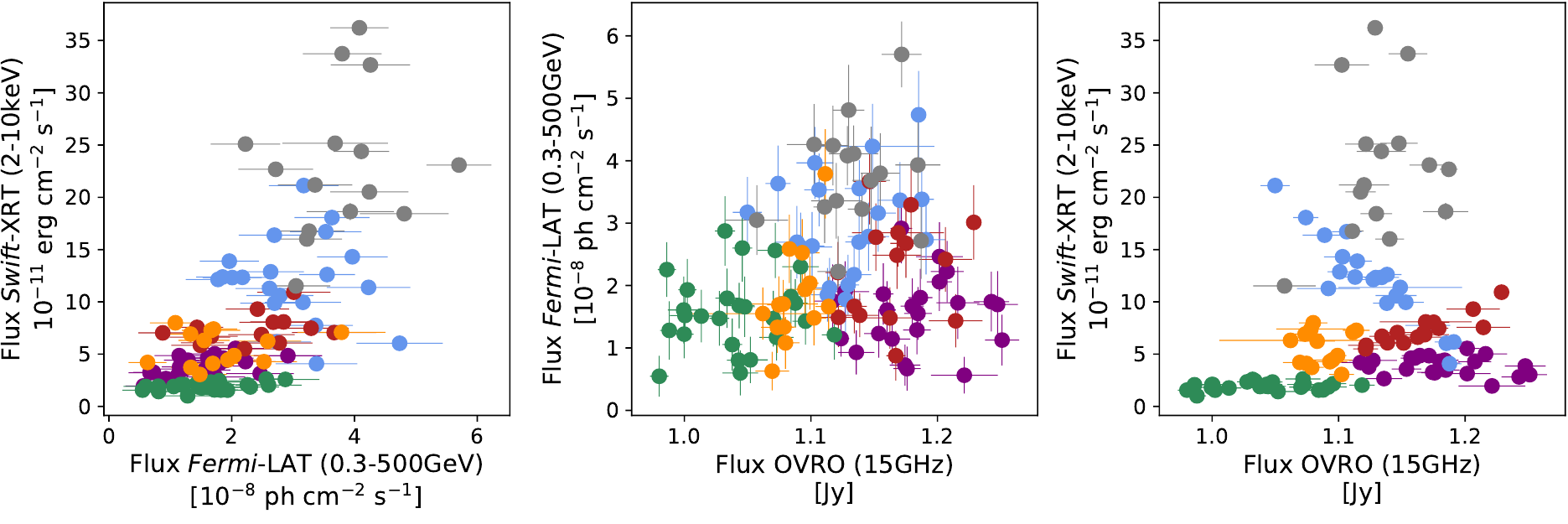}
    \caption{Correlations between different input fluxes with the same color schemes as in a).}
    \label{fig:Corr_of_cl6}
      \vspace*{-30ex}  % Tune this to the image height.
      \hspace{-3cm} \textcolor{lightgray}{\rotatebox{0}{\scalebox{1.5}{PRELIMINARY}}}
    \vspace*{26ex}
\end{subfigure}
\caption{Obtained realization of clusters using the AIC with an optimized number of 6 clusters indicated by the colors.}
\label{fig1}

\end{figure}

Figure~\ref{fig:lc_cl_of_cl6} shows the cluster assignment along the MWL LC together with the obtained models among the data distributions. For comparison, the low state identified in \cite{2023ApJS..266...37A} is shown, which is also identified by the clustering algorithm with only minor changes in the time interval. However, the clustering additional identifies different flaring and intermediate emission states distributed among different time intervals. 

To investigate what is driving the clustering, we checked the correlations between the different input fluxes as shown in Figure~\ref{fig:Corr_of_cl6}. As already indicated in Figure\,\ref{fig:lc_cl_of_cl6}, the clusters show quite a clear division by increasing flux level for the X-rays and, on a slightly less pronounced level, for the $\gamma$-rays. This is supported by the correlated behavior seen when comparing the X-ray to the $\gamma$-ray regime (left panel of Figure~\ref{fig:Corr_of_cl6}).
While we do not learn more from the cluster patterns in the two-dimensional correlation when considering the X-ray and $\gamma$-ray data compared with their one-dimensional LC cluster distributions, the radio band data can be better understood using the two-dimensional comparison. Comparing the radio band fluxes with the X-ray regime (right panel of Figure~\ref{fig:Corr_of_cl6}), it is indicated that for different flux levels in X-rays, the clusters are divided in low and high fluxes in the radio band, especially at the lower X-ray fluxes. However, cluster 1 (blue) seems to go against any of the other trends. 

\section{Discussion}

The six clusters identified above can be interpreted as six emission states of Mrk\,501 most dominantly divided by the X-ray flux, where Cluster 3 (grey) (see Figure\,\ref{fig1}) can be interpreted as a flaring state, Cluster 4 (green) as low state and the other clusters as different intermediate states. The low state identified by this novel method coincides well with its original identification using the Bayesian block method on the VHE LC \cite{2023ApJS..266...37A}. The X-ray regime is known for being one of the most variable wavebands of Mrk\,501 \cite{Mrk501_MAGIC_2013,2023ApJS..266...37A}, and therefore it is not surprising that it dominates the clustering. The high-energy $\gamma$-rays follow the same flux level trends among the clusters as the X-rays, however, less pronounced. This can be explained with the lower degree of variability and sensitivity in this waveband compared to the X-rays. The similar behavior between the wavebands is expected because their long-term LCs have previously shown a significant correlation without time lag \cite{2023ApJS..266...37A}.

Another driver of the clustering is the radio flux level, although with a dependence on the X-ray fluxes as shown by their correlation (right panel of Figure\,\ref{fig:Corr_of_cl6}). For the lower X-ray flux levels, the radio flux is divided in high and low fluxes by the clustering. The only exception is Cluster 1 (blue), which goes against this trend. This different behaviour of Cluster 1 is, however, not seen when we take into account the $\sim$100\,days time lag that was found between OVRO and \textit{Fermi}-LAT in \cite{2023ApJS..266...37A}. For this case, the division between the X-ray and radio fluxes is distinctive without any opposing trends.

The different behavior in the X-ray and radio regime suggests that multiple zones maybe responsible for the different wavebands. The high-energy component of the emission might only contribute partly to the radio emission, while another emission zone explains the rest of the radio emission. Variable behavior of the second zone could then explain the division in radio flux levels, which is not proportional to the X-ray flux but shows up as different clusters. It is nowadays assumed that the radio emission originates partly from more extended regions than the high-energy component (see e.g. \cite{Mrk501_MAGIC_2014}). With further studies and additional data and wavebands, the clustering could offer the potential to help identifying which part of the radio emission is associated with which emission zone.

\section{Summary \& conclusions}
For the first time, we used a clustering algorithm to identify emission states of a blazar. This novel approach can take into account multiple wavebands simultaneously, compared to conventional methods, such as Bayesian blocks \cite{Scargle_2013}, using only single LCs, and is independent of the order of time of the data points. This means, it can also group together data points which are separated in time by other emission states and allow to find separated occurrences of the same emission states.

Our first feasibility study allowed us to identify six emission states of Mrk\,501, which are in agreement with previous results, such as the dominating X-ray variability or the historical two-year long low-activity state of Mrk\,501 \cite{2023ApJS..266...37A}. However, it additionally connects intermediate and flaring states at different time epochs and reveals a distinctive pattern between the clustering in X-ray and radio flux indicating multiple emission zones. 

Further studies will need to be conducted to investigate if the identified clusters actually represent different emission states of the blazar. However, if proven successful, the clustering can, on the one hand, be used to group certain emission properties for a single object enabling time evolution studies of long-term LCs and easing theoretical modeling. On the other hand, with the growing monitoring data sets, cluster algorithms might offer a possibility to identify common emission states among various blazars and even identify new (sub-)populations.

\bibliographystyle{aasjournal}
\bibliography{bib}

\begin{acknowledgments}
This research has made use of data from the OVRO 40-m monitoring program \cite{Richards_2011} supported by private funding from the California Insitute of Technology and the Max Planck Institute for Radio Astronomy, and by NASA grants NNX08AW31G, NNX11A043G, and NNX14AQ89G and NSF grants
AST-0808050 and AST- 1109911.
 \end{acknowledgments}

% \begin{thebibliography}{99}
% \bibitem{Jain_1988}
% A. K. Jain \& R. C. Dubes, Prentice-Hall Inc. (1988)
% \bibitem{Cluster_Springer}
% D. Xu \& Y. Tian, Annals of Data Science 2.2, pp. 165–193 (2015) 
% \bibitem{kmeans}
% J. Macqueen, 5-th Berkeley Symposium on Mathematical Statistics and Probability, pp. 281–297 (1967)
% \bibitem{GMM}
% C. Rasmussen, Advances in Neural Information Processing Systems, Vol. 12. MIT Press (1999)
% \bibitem{DBScan}
% M. Ester et al., Proceedings of the Second International Conference on Knowledge Discovery and Data Mining, AAAI Press, pp. 226–231 (1996)
% \end{thebibliography}
%% Full authors list (ONLY FOR COLLABORATIONS)
%\clearpage
%\section*{Full Authors List: \Coll\ Collaboration}
%
%\noindent \textbf{Note comment afterwards:} Collaborations have the possibility to provide an authors list in xml format which will be used while generating the DOI entries making the full authors list searchable in databases like Inspire HEP. \\
%
%\scriptsize
%\noindent
%first.author$^1$, 
%second.author$^2$, 
%third.author$^3$ % .... more names
%and 
%last.author$^{n}$ \\
%
%\noindent
%$^1$first.affiliation.
%$^2$second.affiliation. % .... more affiliation
%$^{m}$last.affiliation.

\end{document}